\documentclass[useAMS,usenatbib]{mn2e}
\usepackage{graphicx}
\title{External Sources of Poynting Flux in MHD Simulations of Black Hole Ergospheres}
\author[Brian Punsly]{Brian Punsly \\
4014 Emerald Street No.116, Torrance CA, USA 90503 and \\
International Center for Relativistic Astrophysics,
I.C.R.A.,University of Rome La Sapienza, I-00185 Roma, Italy\\
E-mail: brian.m.punsly@L-3com.com or brian.punsly@gte.net}
\begin{document}
\maketitle \label{firstpage}
\begin{abstract}This article investigates the physics that is responsible for
creating the outgoing Poynting flux emanating from the ergosphere
of a rotating black hole in the limit that the magnetic energy
density greatly exceeds the plasma rest mass density (magnetically
dominated limit). The underlying physics is derived from published
3-D simulations that obey the general relativistic equations of
perfect magnetohydrodynamics (MHD). It is found that the majority
of the outgoing radial Poynting flux emitted from the magnetically
dominated regions of the ergosphere is injected into the
magnetosphere by a source outside of the event horizon. It is
concluded that the primary source of the Poynting flux is
associated with inertial forces in the magnetically dominated
region proper or in the lateral boundaries of the region. However,
the existing numerical data does not rule out the possibility that
large computational errors are actually the primary source of the
Poynting flux.
\end{abstract}
\begin{keywords}
black hole physics -- methods:numerical.
\end{keywords}
There are two known theoretical mechanism for producing
field-aligned outgoing poloidal Poynting flux, $S^{P}$, at the
expense of the rotational energy of a black hole in a
magnetosphere that is magnetically dominated. There are
electrodynamic processes collectively called Blandford-Znajek
mechanisms in which currents flow virtually parallel to the proper
magnetic field direction (force-free currents) throughout the
magnetically dominated zone all the way to the event horizon
\cite{blz77,phi83,thp86}. Therefore, these electrodynamic currents
have no source within the magnetically dominated black hole
magnetosphere. Alternatively, there is the GHM
(gravitohydromagnetic) dynamo in which large relativistic inertia
is imparted to the tenuous plasma by black hole gravity that in
turn creates a region of strong cross-field currents (inertial
currents), $J^{\perp}$, that provide the source of the
field-aligned poloidal currents, $J^{P}$ supporting $S^{P}$ in an
essentially force-free outgoing wind (see \cite{pun01} or the
seminal articles \cite{pun90,pun91}). The force-free
electrodynamic current flow is defined in terms of the Faraday
tensor and the current density as $F^{\mu\nu}J_{\nu}=0$. As a
consequence, $\mathbf{J}\cdot\mathbf{E}=0$, so $S^{P}$ in a
force-free magnetosphere must be injected from a boundary surface.
The two types of sources associated with these two types of
Poynting fluxes are quite distinct in the ergopshere: $J^{\perp}$
provides the $\mathbf{J}\cdot\mathbf{E}$ source in Poynting's
Theorem and the force-free (electrodynamic) component of $S^{P}$
emerges from a boundary source at the event horizon.
\par Numerical models can be useful tools for understanding the source of $S^{P}$
emerging from the ergosphere of a black hole magnetosphere. Some
recent 3-D simulations in \cite{dev03,hir04,dev05,dev06,kro05}
show $S^{P}$ emanating from magnetically dominated funnels inside
of the vortices of thick accretion flows. In these simulations,
$J^{\perp}\approx J^{\theta}$(in Boyer-Lindquist coordinates
which are used throughout the following) inside of the ergosphere,
since the poloidal field settles to a nearly radial configuration
early on in the simulation \cite{hir04}. In principle, one can
clearly distinguish the amount of $S^{P}$ emerging from the
ergosphere in a simulation that is of electrodynamic origin (as
proposed in \cite{blz77}) from the amount due inertial effects
(the GHM theory of {\cite{pun01}) by quantifying the relative
strengths of $S^{P}$ emerging from the inner boundary (the
asymptotic space-time near the event horizon) with the amount
created by sources within the ergosphere. In the high spin rate
simulations in questions, over 72\% of $S^{P}$ emerging from the
ergopsheric funnel is created outside of the inner boundary.
\par The simulation that is analyzed, "KDE" is one of a family of
solutions described in \cite{dev03,hir04,dev05,dev06,kro05}. The
solutions within the family can be distinguished by their spin
rate as defined by the ratio of black hole angular momentum to
mass, $a/M$. The KDE simulation spins the most rapidly with,
$a/M=0.998$. All of the simulations begin with a torus surrounding
the black hole which would be stable if not for the ad hoc
introduction of loops of poloidal magnetic field along the equal
pressure contours. This destabilizes the torus as the shearing of
the loops in the differentially rotating plasma creates magnetic
torques on the plasma that initiates an accretion flow. The
angular momentum removal by the field is sustained by the inward
flow of gas that approaches a centrifugal barrier near the black
hole. This barrier creates an "inner edge" of the accretion flow
that forms a funnel roughly along the gravitational equipotential
surface. The subsequent accretion of gas is restricted primarily
to the equatorial plane. As the magnetic flux loops accrete, the
upper portion gets stretched vertically by gas pressure gradients
and electromagnetic forces forces away from the hole and the
inward part of the loop gets severely twisted azimuthally as it
approaches the horizon. The field lines become inextricably
tangled. The net result is the formation of magnetic field lines
that are approximately radial, but highly twisted azimuthally, in
the region near the black hole, restricted to the funnel. \par The
KDE solution with $a/M=0.998$ has by far the most powerful outward
Poynting flux in the funnel within in the family of simulations.
It is the only solution in which this energy flux is comparable
with the energy flux emerging in a mildly relativistic jet created
within the funnel wall (the boundary or shear layer between
accretion disk and the funnel). The KDE simulation is the only one
that appears capable of supplying the high power relativistic jets
that are observed in kinetically dominated FRII quasars (the
kinetic luminosity equals or exceeds the bolometric luminosity of
the accretion flow). As an example in Cygnus A and 3C 82, the jet
kinetic luminosity is roughly 5-10 times the bolometric luminosity
of the thermal emission from the accretion flow (see the
supplementary material in \cite{sem04}). Consequently, from an
astrophysical perspective the KDE simulation is very intriguing.
\par Ostensibly, one might think that the spin rate of $a/M=0.998$
is a benign change from the KDP simulation in
\cite{dev03,hir04,dev05,dev06,kro05} with $a/M=0.9$. A significant
difference is that the surface area of the equatorial plane in the
the ergopshere (the active region of the black hole) is almost 2.5
times larger in the $a/M=0.998$ case \cite{pun01,sem04}.
\par The paper is organized as follows. Section 1 introduces
notation and basic concepts that describe MHD black hole
magnetospheres. This is not new material, but it is taken from
existing literature and provides a framework for the discussion to
follow. The second section presents Poynting's theorem in curved
space-time in integral form, the energy conservation equation for
the electromagnetic field. The primary result of this section
appears in section 2.3: over 72\% of $S^{P}$ emerging from the
ergopsheric funnel during the course of the simulation is created
outside of the inner boundary. This result follows directly from
the data presented in \citet{kro05} and does not require any
interpretation or manipulation of the data. In order to verify
that the data sampled is faithful to the simulation in
\cite{kro05}, we note an independent result from the same work in
section 2.4: there is also a strong source of field aligned
poloidal angular momentum flux, $S_{L}^{P}$ in the ergosphere. The
source of $S_{L}^{P}$ is co-spatial with the source of $S^{P}$ and
of similar magnitude, i.e., over 70\% of $S_{L}^{P}$ that emerges
from the ergosphere is created external to the event horizon. This
means that either the simulation truly has an electromagnetic
source in the ergosphere or there are two independent plots in
\citet{kro05} that have coincidental errors. In section 3, it is
shown that there is a unique physical explanation of the source of
$S^{P}$ and $S_{L}^{P}$ that is consistent with Faraday's law
averaged over time and azimuth (the sampled data from the
simulation is averaged over time and azimuth in \citet{kro05}). A
physically acceptable source must be a cross-field poloidal
current (inertial current, since it is associated with strong
$\mathbf{J}\times\mathbf{B}$ forces in the momentum equation of
the plasma). By Ampere's law this current supports a toroidal
magnetic field that is the field component required for a jump in
both $S^{P}$ and $S_{L}^{P}$ (averaged over time and azimuth) in
the ergosphere. The GHM theory of black hole magnetospheres is
based on such a current. The only thing that would invalidate this
straightforward calculation are significant numerical errors in
the simulation. Section 3 is the only section in this paper that
manipulates the data in \citet{kro05} and so it does not carry the
same weight as the direct results of section 2. However, the
analysis of section 3 is useful (even if it turns out to be
misleading) for the development of future simulations because such
an outcome would indicate gross inconsistencies within the data in
\citet{kro05}. Section 4 compares the results seen in KDE with
other simulations of rotating black hole magnetospheres. This
motivates a discussion in the section 5 of the various numerical
techniques and why they tend to either support or not support the
KDE simulation, in particular the differences with the simulations
in \cite{gam04,mck05} are highlighted. Possible sources of
numerical error are explored in terms of computational technique
and coordinate systems. There is also a discussion of the need for
3-D simulations versus 2-D simulations. In the conclusion, we
stress the need for further simulations to see if this is indeed a
valid interpretation of the data. Some suggestions on how future
data should be sampled are offered. Key parameters to monitor are
mentioned and likely candidates for numerical artifacts are noted
as a guide to help future efforts sort out this complicated
turbulent problem.
\section{Physical Quantities in Boyer-Lindquist Coordinates}
The Kerr metric (that of a rotating uncharged black hole),
$g_{\mu\nu}$, in Boyer-Lindquist coordinates $(r,\theta,\phi,t)$,
is given by a line element that is parameterized by the black hole
mass, $M$, and the angular momentum per unit mass, $a$, in
geometrized units \cite{thp86}. We use the standard definitions,
$\rho^{2}=r^{2}+a^{2}\cos^{2}\theta$ and $\Delta =
r^{2}-2Mr+a^{2}$, where $\Delta=0$ at the event horizon,
$r_{_{+}}=M+\sqrt{M^{2}-a^{2}}$. The "active" region of space-time
is the ergosphere, where black hole energy can be extracted,
$r_{_{+}}<r <r_s = M + \sqrt{M^2 - a^2 \cos^2 \theta}$
\cite{pen69}.
\par The flux of electromagnetic angular momentum along the poloidal
magnetic field direction is the component of the stress-energy
tensor, $T_{\phi}^{\; r}= [1/(4\pi)]F_{\phi\alpha}F^{\alpha r}$,
in the approximation that the field is radial. In steady state,
the electromagnetic angular momentum flux per unit poloidal
magnetic flux is the toroidal magnetic field density:
$-B^{T}\equiv\sqrt{-g}\,F^{\theta r}$, where
$g=-\rho^{4}\sin^{2}{\theta}$ \cite{phi83,pun01}. Similarly, the
electromagnetic energy flux along the poloidal magnetic field
direction, $S^{P}$, is the component, $T_{t}^{\; r}=
[1/(4\pi)]F_{t \alpha}F^{\alpha r}$, in the approximation that the
poloidal field is radial \cite{thp86}. In steady state, the energy
flux per unit poloidal magnetic flux is $-(\Omega_{_{F}}/c)B^{T}$,
where $\Omega_{_{F}}$ is the field line angular velocity
\cite{phi83,pun01}. Consequently, $B^{T}$ is useful for
quantifying the energy and angular momentum fluxes as steady state
is approached. From Ampere's law,
\begin{eqnarray}
\sqrt{-g}\frac{4\pi}{c}J^{\theta}=B^{T}_{\;\;
,r}+(\sqrt{-g}F^{\theta \phi})_{,\phi}+(\sqrt{-g}F^{\theta
t})_{,t}\;,
\end{eqnarray}
evaluated at late times (as an approximate steady state is
reached), one expects simulation data that is averaged over
azimuth to obey $\sqrt{-g}J^{\theta}\approx B^{T}_{\;\; ,r}$.
Therefore, at late times, $J^{\theta}$ is a potential source for
the current system that supports $S^{P}$.
\par Even when a system
has not reached a time stationary state, one can introduce a
well-defined notion of $\Omega_{_{F}}$ that becomes the field line
angular velocity in the steady state. If the field is nearly
radial one can simply define the expression, $F_{t\theta}\equiv
 -\Omega_{_{F}}F_{\theta\phi}$. With this definition, $\Omega_{_{F}}$
is a function of space and time and in steady state it becomes a
constant along a perfect MHD flux tube \cite{phi83}. Therefore,
the EMF across the magnetic field is
$-\Omega_{_{F}}F_{\theta\phi}$ by definition.
\section{The Source of Ergospheric Poynting Flux in the KDE Simulation}
The simulation "KDE" of \cite{dev03,hir04,dev05,dev06,kro05} is of
the most interest since it generates an order of magnitude more
$S^{P}$ than any of the other simulations \cite{kro05}. The
magnetically dominated funnel spans the latitudes $0^{\circ} <
\theta < 55^{\circ}$ at the inner calculational boundary, near
$r_{_{+}}$. Quantifying the magnetic dominance in the funnel is
the pure Alfven speed, $U_{A}=B^{P}/(\sqrt{4\pi n \mu}c)$, where
$n$ is the proper number density, $\mu$ is the specific enthalpy
of the plasma and $B^{P}$ is the poloidal field strength: within
the funnel, $10 < U_{A}^{2} < 10^{4}$.
\begin{figure*}
\includegraphics[width=123 mm]{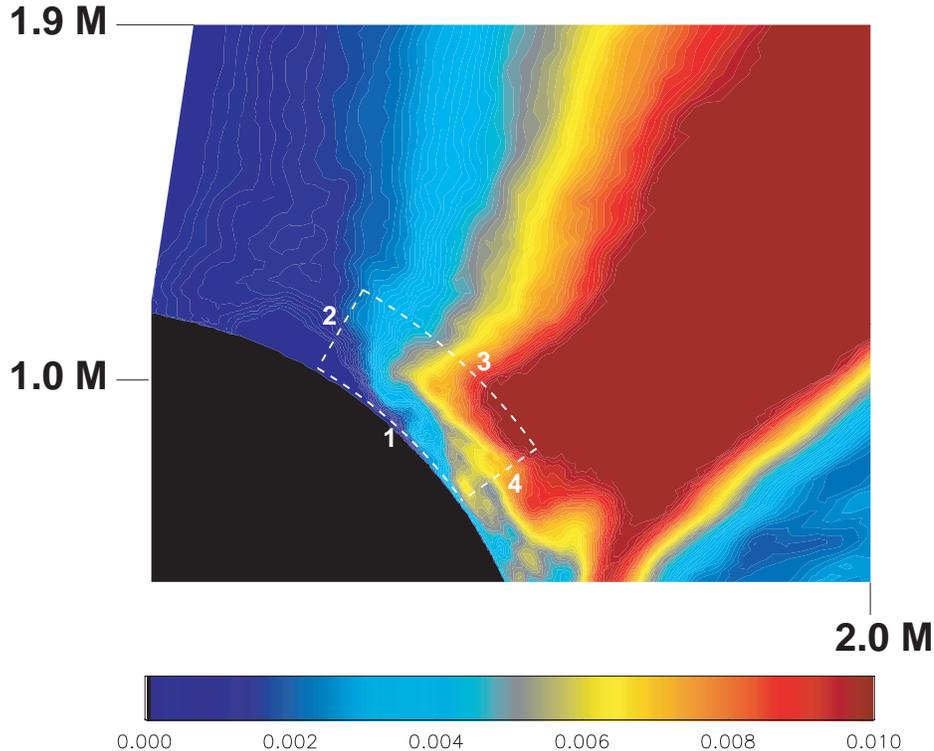}
 \caption{The azimuthally averaged and time averaged (over 75\% of the simulation
    that ends at $t=8080 M$) Poynting flux from the
    model KDE ($a/M=0.998$) of Krolik et al (2005). The figure is a magnification of the
    inner region of figure 10 of Krolik et al (2005). It is an excision of a region,
    $0^{\circ} < \theta < 65^{\circ}, r\sim r_{+}$ that is a little
    larger than the ergospheric portion of the magnetically dominated
    funnel, $0^{\circ} < \theta <55^{\circ}, r\sim r_{+}$. The majority of $S^{P}$ switches-off in a
thin layer near $r=1.3M$
    - $r=1.5M$ (see Krolik et al (2005) for a description of the units on the color bar). A
    Gaussian pillbox, $30^{\circ} < \theta < 55^{\circ}$, is drawn as a dashed white contour for
    use in Poynting's Theorem. There are 26 grid zones between the
    inner boundary, $r=1.175M$ and $r=1.5M$. The plot is provided courtesy of John Hawley. }
    \end{figure*}
\subsection{Poynting's Theorem} The source of
$S^{P}$ can be determined from the integral version of Poynting's
Theorem, which is the integral of $T^{\nu}_{t\;\;
;\nu}=F_{t\nu}J^{\nu}$ combined with Stokes' theorem. In
Boyer-Lindquist coordinates, the curved space-time equivalent of
the "$\mathbf{J}\cdot\mathbf{E}$" source of $S^{P}$ is
$F_{t\alpha}J^{\alpha}$. In steady state, the source of $S^{P}$ is
$-\Omega_{_{F}}F_{\theta\phi}J^{\theta}$ in the approximation of a
radial field. Without the radial approximation, one needs to
define a poloidal field direction, $\mathbf{B^{P}}$, and a
cross-field poloidal EMF, $E_{\perp}\equiv
 -\Omega_{_{F}}B^{P}$ then the source of $S^{P}$ is
$E_{\perp}J^{\perp}$. This can all be setup with great
mathematically complexity (see \cite{pun01} for sample
calculations) and no additional physical insight. The reader
should remember that the poloidal field is not exactly radial in
the following and the simple notion of $J^{\theta}$ is used to
approximate $J^{\perp}$.
\par The funnel is highly
turbulent and even averaging over $\phi$ does not damp the large
field strength variations that are seen in snapshots of this
rotating non-axisymmetric system (DeVilliers privative
communication 2005). Finding a source of Poynting flux in a time
snapshot is futile since the field strength variability manifests
itself as large fluctuations in the Poynting flux on the inner
calculational boundary near the horizon and throughout the
ergosphere (Hawley private communication 2005). Conversely, over
long periods of time the total outflow of energy swamps the local
random turbulent behavior seen in any time snapshot. For this
reason \cite{kro05} time averaged all the data in order to
extricate the persistent physics from the strong turbulent
fluctuations that obscure the physics in individual snapshots. In
order to understand the source, one is interested in where the
total time integrated Poynting flux that radiates from the
ergosphere originates. This information can be obtained by
performing a time integral of the simulation, once the large
transients have died down. The discrete time average and azimuthal
average of $S^{P}$ in the KDE simulation near $r_{_{+}}$ is shown
in figure 1. The time integral of $S^{P}$ can be approximated by
multiplying the discrete time average by the total time. The
contour map in figure 1 indicates a region of strong outgoing
$S^{P}$ in the evacuated funnel, $30^{\circ} < \theta <
55^{\circ}$, $r\sim r_{+}$. It is clear that $S^{P}$ suddenly
diminishes close to $r_{+}$ at $r\approx 1.3M-1.5M$.
\par In order to investigate possible source terms for $S^{P}$, one needs
to be explicit about what is plotted in figure 1. The data was
averaged over $\phi$ and $t$ (denoted by "$<>$"). Whenever a time
lapse of 80 M occurs within the high time resolution simulation,
data was stored for a time "snapshot". There are 76 of these
snapshots that are averaged in figure 1, steps 26 through 101.
Thus, the discrete time average of Poynting's theorem is relevant
to figure 1,
\begin{eqnarray}
&&[2/\pi]\frac{1}{76}\sum_{i=26}^{101}\left[\int
[F_{t\phi}J^{\phi}+F_{tr}J^{r}+F_{t\theta}J^{\theta}]\, dV
\right]_{i}\nonumber\\
&&-[2/\pi]\frac{1}{76}\sum_{i=26}^{101}\left[\frac{d}{dt}\int
T^{\,t}_{t}\, dV\right]_{i}
\nonumber\\
&& =[1/(2\pi^{2})]\frac{1}{76}\sum_{i=26}^{101}\left[\int_{1+3}
\sqrt{-g}F_{t \alpha}F^{\alpha r}\,d\theta\,d\phi\right]_{i}\nonumber\\
&&+[1/(2\pi^{2})]\frac{1}{76}\sum_{i=26}^{101}\left[\int_{2+4}
\sqrt{-g}F_{t \alpha}F^{\alpha \theta}\,dr\,d\phi\right]_{i}\;,
\end{eqnarray}
where the four sides of the Gaussian pillbox of integration in
figure 1 are the four curves labelled "1 -4" and $dV=\sqrt{-g}dr
d\phi d\theta$ is the spatial volume element. Periodic boundary
conditions are imposed at $\phi=0$ and $\phi=\pi/2$. Hence,
$<T_{t}^{\; \phi}>$ is equal at both boundaries and does not
contribute to surface integral on the RHS of (2).
\subsection{Limitations of the Existing Data}The data that was stored in
the run of the KDE simulation does not allow us to compute all of
the terms in Poynting's theorem, (2), in each of the coarse time
steps, "i". For example, since at least three consecutive fine
resolution time steps were not stored at each of the 76 coarse
time steps, one can not compute the time derivative or the current
density (from Ampere's law) that appear in Poynting's theorem.
This highlights the limitation of single snapshot information. A
computation of $T_{t\; ;\nu}^{\,\nu}$ in a single noisy snapshot
never includes the $d/dt(T^{t}_{\;t})$ which is likely to dominate
in a turbulent environment, nor can one compute $J^{\mu}$. Thus,
such a calculation is likely to be very misleading. For this
reason the time averaged data presented in \cite{kro05} is
considerably better. Even though $J^{\mu}$ can not be computed,
one expects that the $d/dt(T_{t}^{\;t})$ term time averages to
zero if the system is physical (i.e., a finite field energy
density in a small spatial region should not decay as a persistent
source of Poynting flux). Even so, we are limited by not being
able to compute $\mathbf{J}\cdot\mathbf{E}$. However, there are
conclusions that can be reached. First of all figure 1 can be used
to compare the total $S^{P}$ radiated from $r_{_{+}}$ and $r_{s}$.
This is the subject of section 2.3 and the results do not depend
on any analysis of the data, but follow directly from the data.
Secondly, we have time averaged information on $<T_{\phi}^{\; r}>$
from \cite{kro05} which allows us to restrict the source terms in
(2). Using this information in section 3.3, we deduced that there
is only one source that is consistent with the $<T_{\phi}^{\; r}>$
and the $<T_{t}^{\; r}>$ data from \cite{kro05},
$<\mathbf{J}\cdot\mathbf{E}>$. However, without having the actual
data, complete with differential information, one can not assess
the level to which numerical errors invalidate the conclusions of
section 3.3.
\subsection{The Horizon is not the Primary Source of
Poynting Flux}The total Poynting flux emanating from inner
ergosphere, $r<1.5M$, after the large transients have died down,
$2000M<t<8080M$, is $\approx (20/\pi)M\sum_{i=26}^{101}\int_{3}
\sqrt{-g}F_{t \alpha}F^{\alpha r}\,d\theta\,d\phi$. Similarly, the
time integrated $S^{P}$ emanating from inner boundary (horizon),
after the large transients have died down, $2000M<t<8080M$, is
$\approx (20/\pi)M\sum_{i=26}^{101}\int_{1} \sqrt{-g}F_{t
\alpha}F^{\alpha r}\,d\theta\,d\phi$. The result of significance
from this plot is that $\sum_{i=26}^{101}\int_{3} \sqrt{-g}F_{t
\alpha}F^{\alpha r}\,d\theta\,d\phi >
3.5\mid\sum_{i=26}^{101}\int_{1} \sqrt{-g}F_{t \alpha}F^{\alpha
r}\,d\theta\,d\phi\mid$, i.e. $\int S^{P}\,dt$ emitted from the
inner boundary (horizon) during the interval, $2000M<t<8080M$, is
much smaller than $\int S^{P}\,dt$ emitted from $r_{s}$ during the
interval, $2000M<t<8080M$. Because of the saturation of the dark
red color in the plotting routine, $S^{P}$ might be even larger
above the switch-off layer than indicated in the contour map.
Thus, we only have a lower bound on the strength of $S^{P}$ above
the switch-off layer and it is likely that more than 72\% of the
total $S^{P}$ is created within this thin layer. At this point, we
can conclude that during the interval $2000M<t<8080M$, less than
$28\%$ of $S^{P}$ that emerges from the ergospheric funnel came
from the inner boundary, the rest was created external to the
boundary.
\subsection{The Ergospheric Source of Angular Momentum Flux}
One could conjecture that without the actual data in hand that it
might be difficult to assess whether the source seen in the false
color and the embedded white contours in figure 1 truly reflects
the simulation. Thus, we are compelled to find supporting data.
First of all, is the external source a plotting routine artifact?
Secondly, one might ask if there is any other supporting
information that there is a strong electromagnetic source external
to the horizon? Fortunately, there are a family of line plots of
the electromagnetic angular momentum flux as a function of
$\theta$ and $r$ in Figure 7 of \cite{kro05}. Each line plot
represents the averaged (in $t$ and $\phi$) electromagnetic
angular momentum flux, \begin{eqnarray}
 &&<T_{\phi}^{\; r}>=
[1/(4\pi)][<F_{\phi t}F^{t r}>+<F_{\phi\theta}F^{\theta r}>]\;,
\end{eqnarray}
as a function of $\theta$ on a $r$=constant surface. Figure 7
shows that $<T_{\phi}^{\; r}>$ is created predominantly in the
ergosphere, outside of the horizon. Over 70\% of the
electromagnetic angular momentum flux in the funnel is created by
sources within the ergopshere for $25^{\circ} < \theta <
55^{\circ}$ in a location that is cospatial with the source of
Poynting flux that is prominent in figure 1. Since there are two
strong electromagnetic sources of similar magnitude in the same
location in two independent plots of the KDE simulation, it is
clear that the numerical data represents an electromagnetic
source. However, the data in \cite{kro05} does not allow one to
determine if the source is physical or if it is a numerical
artifact.
\section{Possible Sources of Poynting Flux}There are
three possible source terms for this increase in $S^{P}$ when (2)
is applied to the pillbox in figure 1. We discuss each of these
potential source terms below.
\subsection{Decay of Field Energy} The second term on the LHS of (2) represents the decay
 of locally stored magnetic field energy $d\epsilon/dt$,
 \begin{eqnarray}
 &&\int\frac{d\,\epsilon}{d\,t}d\,t\approx-80M\sum_{i=26}^{101}\left[\frac{d}{dt}\int
T^{\,t}_{t}\, dV\right]_{i}\;.
\end{eqnarray}
There was no data processed from the KDE run that can assess the
magnitude of this term. In order to get the derivative on the RHS
of (4) in each of the 76 snapshots requires that data be dumped in
three consecutive fine resolution time steps at each of the 76
coarse resolution time steps (clearly 5 consecutive time steps
would be superior). However, this was not done and the simulation
would have to be run again with at least three times as much data
gathered in order to compute this term in (4). Thus, we can not
rule out the significance of this term. The time averaged field
decay from a finite spatial region is not a viable
 physical source of long term $S^{P}$ creation. The existence of a persistent source of
 decaying field strength that does not die off would be an indication that the  simulation
is pathological, spontaneously creating local field energy from
numerical artifacts. This is a serious concern.
\subsection{Latitudinal Poynting Flux}It is possible that Poynting flux flows in the upper and
lower sides, 2 and 4, in figure 1 and gets redirected into a
radial Poynting flux by some unknown mechanism. This latitudinal
Poynting flux, $S^{\perp}$, is defined by the surface terms,
\begin{eqnarray}
&& \int S^{\perp} dt\approx
\frac{20}{\pi}M\sum_{i=26}^{101}\int_{2+4}\sqrt{-g}F_{t
\alpha}F^{\alpha \theta}\,dr\,d\phi\;.
\end{eqnarray} It seems that $S^{\perp}$ would most likely
originate in the funnel wall, $55^{\circ}<\theta<60^{\circ}$ in
figure 1 as opposed to the evacuated polar region of the funnel.
Note that there is a very strong source of $S^{P}$ in the funnel
wall in figure 1. The region is denser than the funnel, $U_{A}\sim
1$, hence this putative source of $S^{\perp}$ would be inertial in
character. The conversion of $S^{\perp}$ from the funnel wall into
$S^{P}$ in a magnetically dominated black hole magnetosphere would
be an entirely new source of Poynting flux that has never been
postulated before in the literature. \subsection{Inertial Current}
The remaining possible source of $S^{P}$ in (2) is the inertial
current term
\begin{eqnarray}&&\int\int\mathbf{J}\cdot\mathbf{E}dVdt\nonumber\\
&&\approx \frac{20}{\pi}M \sum_{i=26}^{101}\left[\int
[F_{t\phi}J^{\phi}+F_{tr}J^{r}+F_{t\theta}J^{\theta}]\, dV
\right]_{i}\;. \end{eqnarray} This is the type of source that
exists in GHM \cite{pun01}. It is shown below that this is the
only source in Poynting's theorem that is consistent with the data
presented in \cite{kro05} if numerical errors are negligible.
\par In order to differentiate between the two possible physical
sources first expand
\begin{eqnarray}
&& T_{t}^{\; r}= [1/(4\pi)][F_{t \phi}F^{\phi
r}+F_{t\theta}F^{\theta r}]\;.
\end{eqnarray} Then average the
coordinate frame Faraday's law over $t$ and $\phi$,
 \begin{eqnarray}
&& <F_{\phi \theta ,t}>+<F_{\theta t ,\phi}>+<F_{t\phi
,\theta}>=0\;,
\end{eqnarray}
\begin{eqnarray}
&& <F_{\phi r ,t}>+<F_{r
t,\phi}>+<F_{t\phi,r}>=0\;.
\end{eqnarray} Notice the simplification that (8) and (9) depend
only on ordinary derivatives because we are working in a
coordinate frame \cite{pun01}. Since $<\mathbf{B}^{P}>$ is
approximately constant and radial in the the funnel for $t>2000M$,
$<F_{\phi \theta ,t}>\approx 0$ and $<F_{\phi r ,t}>\approx 0$ in
(8) and (9), respectively. By the periodic boundary condition on
$\phi$, $<F_{\theta t ,\phi}>= 0$ and $<F_{r t,\phi}>= 0$ in (8)
and (9), respectively. Thus, the averaged Faraday's law in (8) and
(9) reduce to, $<F_{t\phi ,\theta}>\approx 0$ and
$<F_{t\phi,r}>\approx 0$, respectively. These relations can be
simply integrated to give $<F_{\phi t}>\approx 0$ or $\mid
<F_{\phi \theta}>\mid \gg M\mid <F_{\phi t}>\mid$ . Physically, as
in flat space-time rotating axisymmetric magnetospheres, $F_{\phi
t}$ is associated with changes in the poloidal magnetic flux.
\par If $<F_{\phi t}>$ is negligible, as
suggested by Faraday's law, then the time averaged (7) becomes
\begin{eqnarray}
<T_{t}^{\; r}>\approx \frac{1}{4\pi}<F_{t\theta}F^{\theta r}> \;,
\end{eqnarray} in its application to figure 1. In order to
isolate the most prominent contributor to the RHS of (10), to the
change in $<S^{P}>$ at $r\approx 1.4M$, recall the angular
momentum source in Figure 7 of \cite{kro05}. If $<F_{\phi t}>$ is
negligible, as suggested by Faraday's law, then the time averaged
(3) becomes,
\begin{eqnarray}
 &&<T_{\phi}^{\; r}>\approx
[1/(4\pi)][<F_{\phi\theta}F^{\theta r}>]\;,
\end{eqnarray} Since
$<\mathbf{B}^{P}>\approx<F_{\theta\phi}>/\sqrt{g_{\phi\phi}g_{\theta\theta}}$
has only mild variation associated with the geometrical factors
arising from $<\mathbf{\nabla}\cdot\mathbf{B}=0>$, the major
contributor to the source of $S^{P}_{L}$ in the ergosphere must be
a jump in $<F^{\theta r}>$. Therefore, the only physically
consistent reconciliation of the increase in both $<S^{P}>$ and
$<T_{\phi}^{\; r}>$ across the region $1.3M < r<1.5M$ with (10)
and (11) is that the primary source for both is a jump in
$-B^{T}$. Ampere's law, (1), averaged over $t$ and $\phi$
indicates that $J^{\theta}$ is the source of jump in $-B^{T}$ and
therefore the ultimate source of both $T_{\phi}^{\; r}$ and
$S^{P}$ in the ergosphere. In particular, the dominant source term
for $<S^{P}>$ in (2) is $<F_{t\alpha}J^{\alpha}>\approx
<F_{t\theta}J^{\theta}>$.
\section{Comparison With Other Simulations}
In this section, a comparison is made between the ergospheric
electromagnetic source of KDE and dynamics of the ergopshere that
are seen in other simulations of rotating black holes.
\subsection{3-D Simulations} The only other 3-D MHD simulation
that has been performed about a rotating black hole is in
\cite{sem04}. This solution is flawed in that it assumes a
background pressure distribution. The simulation follows the
accretion of an individual flux tube on this background. There is
no back reaction on the pressure distribution. In this limit, it
is shown that the relativistic MHD equations for individual flux
tubes reduce to a nonlinear string equation. However, to higher
order, the trans-field momentum equation is ignored, so the
results are not self consistent. The simulations are more accurate
in the scenario developed in \cite{spr05} in which patches of
nearly vertical magnetic flux accrete sporadically into to an
existing central bundle of strong flux. On the positive side,
these simulations provide much higher resolution than any other
simulation, as thousands of points are used on each flux tube in
the ergosphere. This was argued in \cite{sem04} to be required to
adequately treat the large gradients in the highly twisted and
stretched field lines. It is therefore the only long term
simulation in which numerical diffusion does not cause the field
lines to reconnect in the equatorial plane. The artificial
reconnection seen in other simulations changes the topology of the
field and totally changes the large scale torques in the system as
compressive MHD waves can no longer be launched from the
potentially powerful compressive piston in the equatorial plane of
the ergopshere. This phenomenon produces the most powerful jets
from black holes (via a GHM process) in purely analytical
treatments \cite{pun01,pun90}.
\par Another positive of the \cite{sem04} simulations is that the
flux tube evolution is very clearly displayed. As the flux tube
accretes inward it is pulled and stretched toward the black hole
and becomes extremely twisted. In the ergosphere, a GHM dynamo
(Poynting flux and jet source) forms close to the equatorial
plane. As the flux tube accretes further, the dynamo starts
drifting upward and settles at high latitudes just outside the
horizon as seen in the closeup movies of \cite{sem04}. The late
time location of the dynamo is approximately in the same location
as the Poynting flux source in figure 1.
\subsection{2-D Simulations} We review the numerous 2-D
simulations starting from the most primitive and ending with the
most physically realistic, \cite{gam04,mck05}.
\subsubsection{Time Evolution of the Wald Field} The time
evolution of the Wald vacuum field (this is the curved space-time
equivalent of a uniform magnetic field, i.e., the $l=1, m=0$
moment of the vacuum electromagnetic field, see \cite{pun01})
loaded with plasma has been of particular interest because it
seemed like an astrophysically reasonable poloidal field geometry.
The first simulations of \cite{cak00,koi02,koi03} were short term
due to code and computing limitations. They produce effects very
similar to what is seen in figure 1 and \cite{sem04}. A force-free
electrodynamic (a very tenuous MHD plasma) simulation of the same
problem in \cite{kom04} showed a time stationary state that was
similar to the above for flux tubes that thread the equatorial
plane and a Blandford-Znajek type solution on the field lines that
thread the horizon. The field lines that thread the equatorial
plane drove an electrodynamic jet by a process very similar to the
original GHM ergospheric disk model \cite{pun90}.
\par Subsequently, a very important MHD simulation was performed
on the same field configuration in \cite{kom05} that has altered
our understanding of this problem. The initial state in
\cite{kom05} is a tenuous plasma accreting from the outer
boundaries of the simulation grid. At first, the simulation
resembled \cite{cak00,koi02,koi03} and looked like it might
approach \cite{kom04}. However, as mass accumulated in the
equatorial plane, by nearly vertical accretion, the density went
up so as to violate the force-free approximation used in
\cite{kom04}. The radial gravitational attraction on the dense
plasma stretched and dragged the once vertical field lines in the
equatorial plane radially inward, toward the horizon. The
stretched field lines reconnect in the equatorial plane
essentially creating a local split monopole near the horizon. The
final state is an accretion solution on all the field lines that
thread the horizon, there is no outflow or jet that forms.
\par One should note that this result is largely dependent on the
plasma injection mechanism. In the original model of \cite{pun90}
the plasma is created as positron-electron pairs from a gamma ray
cloud as proposed in \cite{phi83}. As the pairs settle in the
equatorial plane they annihilate in a thin disk. The density never
gets high enough to drag the field lines inward as was seen in the
accretion solution of \cite{kom05} and the solution of
\cite{kom04} is more physically representative in this scenario.
Another interesting initial state is the deposition of magnetic
flux bundles in the ergosphere that accrete differentially to the
plasma as a distinct phase of a two phase accretion flow
\cite{spr05}. In this scenario, the flux tubes are injected into
the ergosphere almost vertically and already possess there own
magneto-centrifugal outflow. The density in the equatorial plane
is therefore low and an accretion type solution as in \cite{kom05}
is not likely to occur. See the introductory remarks to section
5.3 for a discussion of the simulation in \cite{dev07} that shows
some of these expected effects.
\subsubsection{2-D Accretion from Tori} The initial state in
\cite{gam04,mck05} is similar to the torus threaded with magnetic
loops described in the introduction for the simulations in
\cite{dev03,hir04,dev05,dev06,kro05}. There are significant
differences, the simulation is 2-D and \cite{gam04,mck05} use
ingoing Kerr-Schild coordinates instead of Boyer-Lindquist
coordinates. A different numerical code is used as well (see
\cite{gam04,dev03} and references therein). In spite of these
differences, the late time states are very similar, both evolve to
a magnetically dominated funnel threaded by an approximately
radial field with outgoing Poynting flux. There is no large scale
vertical magnetic flux in the equatorial plane of the ergosphere.
Both have a powerful inertial outflow along the funnel wall. This
feature is never described in much detail in \cite{gam04,mck05},
so it is hard to compare the funnel wall jet between the two sets
of simulations, i.e., how much of the outgoing energy budget is in
the funnel and how much from the funnel wall. The main result of
interest here is that the simulations indicate that the
force-free, Blandford-Znajek mechanism drives an outgoing Poynting
flux on field lines in the funnel \cite{gam04}.
\par The most relevant simulation from \cite{mck05} is when
$a/M=0.999$. A significant increase in the Poynting flux from the
funnel, a factor of $5-6$, is seen compared to the $a/M=0.9$
simulation. Not as dramatic as the increase seen in \cite{kro05}.
A check of the Poynting flux contours has revealed no significant
external source of Poynting flux in the ergospheric funnel in
contrast to the KDE simulation (J. McKinney private communication
2005). The implication is that the discrepancy arises either from
the 2-D versus 3-D nature of the simulations, the numerical code
choice or the the coordinate system chosen. No matter what has
caused this discrepancy, the major point of interest is whether
this is a physical difference (i.e., 2-D versus 3-D) or a
numerical error (coordinates choice or code inadequacies).
\section{Possible Sources of the Discrepancy} It was noted above
that the simulations of \cite{gam04,mck05} do not support the data
presented in figure 1 for the KDE simulation. Possible sources of
this discrepancy are discussed below as a guide to help future
numerical efforts that could resolve this important physical
point. The emphasis is on possible sources of numerical errors.
\subsection{Numerical Method} In section 3.1, the creation of
Poynting flux resulting from the decay of field energy could not
be ruled out as possible persistent source. Although this is not
physical, it could result from numerical artifacts. The method
used in \cite{dev03,hir04,dev05,dev06,kro05} is non-conservative.
The field is conserved in the sense that $\nabla\cdot\mathbf{B}=0$
is guaranteed by the numerical method and is not a subsidiary
condition that is imposed. However, global energy is not conserved
in each time step. This can lead to unphysical jumps in the fluid
4-velocity which could create unphysical changes in the field
strength through the frozen-in condition, $F^{\mu\nu}u_{\nu}=0$.
This means that electromagnetic energy could potentially appear
and disappear for no physically or mathematically consistent
reason. In regions of fast oscillations and large spatial
discontinuities such a code always generates extra oscillations
\cite{dev01}. The extra oscillations can allow the field to gain
or lose additional amounts of energy and energy flux. If this
effect is significant there is no obvious apparent reason why it
should preferentially create energy or destroy energy in a time
average. Thus, in of itself, it would not explain the source in
figure 1, unless it is coupled to another numerical problem (see
section 5.2 for example). By contrast the reader should note that
the numerical methods of \cite{kom04,gam04,mck05} are based on a
conservative approach which is therefore much less prone to this
type of error \cite{gam03}.
\par The radial resolution of 26 grid zones between the inner
boundary and the Poynting flux source in figure 1 is probably
adequate. Although, since this feature is suspect more grid zones
would be preferably in future simulations. More of a concern is
the $\theta$ resolution. The grid zones were concentrated near the
equatorial plane since the original intent of the simulations in
\cite{dev03,hir04,dev05,dev06,kro05} was to model the accretion
disk with high resolution. There are 192 zones in the range
$0<\theta<\pi$ and 26 zones inside the Gaussian pillbox of figure
1. Thus, the resolution is not adequate to resolve latitudinal
gradients which could be very important for understanding the
differential version of Poynting's theorem. Furthermore, it is
important to adequately resolve the funnel jet wall since this is
a highly dynamic region that carries most of the outflow energy,
yet it is just a thin layer. Poor resolution of the funnel wall
jet could allow energy and mass to diffuse across the field lines
into the funnel in figure 1.
\subsection{Coordinates} Another concern is the choice of
Boyer-Lindquist coordinates. The Boyer-Lindquist coordinates have
a coordinate singularity at the event horizon. This singularity
could be excised by an inner boundary as in
\cite{dev03,hir04,dev05,dev06,kro05}, but one still has steep
gradients in the metric coefficients outside the horizon. This has
long been recognized as a problem for numerical work by gravity
wave theorists. There is a tendency for strong reflections to be
generated at the inner boundary \cite{cam01}. The subluminal fast
MHD wave speed might be a possible distinction from gravity wave
physics that allows Boyer-Lindquist coordinates to be used
successfully in MHD. The intention was to place the boundary so
close to the horizon that all information would be flowing inward
and reflections would not be an issue. For this assumption to
hold, the flow must be going inward super-magnetosonic (faster
than the fast MHD wave speed) outside of the inner boundary.
However, the super-fast nature of the flow was never demonstrated
for KDE.  A rough order of magnitude calculation in \cite{pun01}
yields the lapse function, $\alpha$ at the fast point,
$\alpha\equiv \sqrt{\Delta}\sin{\theta}/\sqrt{g_{\phi\phi}}\approx
U_{A}^{-1}$. This puts the fast point at $r\approx 1.0651M$, at
$\theta=30^{\circ}$ for example. This is inside of the inner
calculational boundary at $r=1.175M$. The interaction of the flow
with the boundary is another potential artificial numerical source
of Poynting flux.
\par In order to resolve this problem, gravity wave researchers
have gone to Kerr-Schild coordinates \cite{cam01}. There are no
large gradients in the metric near the horizon or coordinate
singularity. An inner computational boundary is placed inside the
horizon. This method has been employed by
\cite{kom04,gam04,mck05}. The only way that information can get
from inside or near the horizon and react back on the upstream
flow is by numerical diffusion. \subsection{2-D versus 3-D} The
choice of 3-D by \cite{dev03,hir04,dev05,dev06,kro05} is certainly
preferential to 2-D for physical reasons. For, example flux tube
evolution is vastly different if interchange instabilities are
allowed. These are impossible in 2-D. Unfortunately, 3-D is much
more cumbersome numerically. This is the reason that the KDE
simulation was not sampled adequately to determine the 4-current
density. As of today, the numerical code of \cite{dev01} is the
only code that is efficient enough to produce long term
self-consistent 3-D MHD simulations. One should question how
accurate the 2-D expedience is at reproducing the physics of 3-D
time evolution in black hole magnetospheres. The external source
in figure 1 might be the first indication that 3-D is required to
capture all of the physical effects.
\par The most intriguing question is whether
there is a theoretical reason why 2-D can yield different results
than 3-D simulations. For example, the original ergospheric disk
model of \cite{pun90} is axisymmetric macroscopically, but
microscopically it is 3-D. Buoyant flux tubes are created by
reconnection at the inner edge of the ergospheric disk and recycle
back out into the outer ergosphere by interchange instabilities.
The details of this model could not exist in 2-D. The 3-D
requirement to model the physics of vertical flux in the
equatorial plane of the ergopshere is suggested by new unpublished
simulations in \cite{dev07}. De Villiers has placed a Wald field
(as in section 4.2.1) inside of the torus in the initial state,
the simulation is otherwise similar to the KDP simulation with
$a/M=0.9$ in \cite{dev03,hir04,dev05,dev06,kro05}. A weak Wald
field gets dragged inward with very little modification to what
was found for KDP in agreement with \cite{kom05}. A strong Wald
field completely disrupts the torus. A moderate strength field can
be chosen that is sufficient to impede the accretion flow, but not
disrupt the disk. As accreting flux builds up in the ergosphere,
accretion gets arrested and large episodic releases of jet energy
occur until the built up magnetic pressure gets relieved
(presumably by interchange instabilities). The recycling magnetic
pressure resembles the recycling of magnetic flux tubes in the
ergospheric disk model \cite{pun90}. The cycle repeats on time
scales on the order of the Alfven wave crossing time for the
ergopshere. This is quite a contrast to the 2-D simulation of the
Wald field in \cite{kom05}. In the 2-D scenario, the vertical flux
gets overwhelmed by plasma inertia and dragged into the hole. In
the 3-D simulation, the vertical flux impedes the accretion flow
and there is significant vertical flux that is not pinned inside
the hole as evidenced by the recycling magnetic pressure in the
ergopshere.
\par A more general motivation for 3-D is indicated by the contrast
in the field configurations that give rise to either
electrodynamic or GHM energy generation in the existing simulation
literature.
\subsubsection{Electrodynamic Field Configurations} The ultimate
distinction between the electrodynamic solutions and the GHM
solutions is whether the magnetic field is organized in the
ergosphere so that a tenuous plasma can flow approximately
force-free (the electrodynamic condition). For example, this can
occur if a weak field threading a dense plasma (i.e., $U_{A}<1$)
is deposited near the event horizon by an accretion flow. In this
scenario, the field lines are dragged inward radially and twisted
up azimuthally by plasma inertia. As plasma is depleted by
accretion inward and magneto-centrifugal and thermal outflow, the
flux tube has been essentially pre-configured (twisted and
stretched) by an initial nonforce-free, inertially dominated
transient, so that subsequent tenuous accretion can proceed
approximately force-free. This is the manner in which the
magnetopshere is built up in
\cite{dev03,hir04,dev05,dev06,kro05,gam04,mck05}.
\par Alternatively, a strong magnetic field can exist near the
black hole and can be almost vertical, as in
\cite{cak00,koi02,koi03,sem04,kom04} and the initial state in
\cite{kom05}. The tenuous plasma is restricted from a force-free
inflow. However, if an inertially dominated transient state
(nonforce-free) can form as in the dense equatorial condensate of
\cite{kom05}, this will drag the field lines inward radially
towards the horizon and twist them azimuthally in the process. The
field has then been prepared for future approximately force-free
(electrodynamic) inflow in the ergopshere.
\par An initial inertially dominated transient appears to be required to
modify an accreted field to a force-free configuration. This is
evidenced by the simulation of \cite{kom04}. Purely electrodynamic
effects could not distort the field lines so that a flux tube that
did not initially allow force-free inflow into the horizon could
ever change. In this limit, in which the plasma never gains any
substantial inertia, the field lines never get dragged inward into
the hole by a transient. There was no evolution to force-free flow
into the black hole on these flux tubes.
\subsubsection{GHM Field Configurations} If the field lines do not
allow an approximately force-free flow into the horizon, the
induced stresses imposed by the plasma creates a GHM dynamo in the
ergopshere, by definition. Thus, the transients described above
are transient GHM dynamos for example. The question posed by KDE
is, are there ever any stable GHM dynamos?
\par The simulation of \cite{kom04} and the analytic model of
\cite{pun90} are examples of stable GHM field configurations. In
these cases a vertical magnetic flux through the equatorial plane
prevents the plasma form flowing towards the horizon. The field
has to be stretched radially, twisted azimuthally and possibly
crossed by the tenuous plasma in order for the flow to proceed
inwards. The struggle between the strong field and the
relativistic inertia imparted to the plasma by the gravitational
field is the source of the GHM dynamo \cite{pun01,sem04}.
\subsubsection{Field Configuration: 2-D versus 3-D Time Evolution} The scenario
envisioned in \cite{spr05} requires 3-D. There needs to be an
extra dimension to allow the degree of freedom that permits
evolution of the magnetic flux tubes somewhat independent of the
mass accretion. The flux tubes need interchange instabilities in
order to allow plasma to accrete around them (i.e., so they can
"swim" in the accretion flow). In 2-D ideal MHD, the accreting
plasma will always build up in density until it can push the field
lines ahead of it by ram pressure or just drag the field lines
threading it into the horizon. This process will continue until
the magnetic pressure roughly equals the ingoing ram pressure. The
plasma can never go around a flux tube. Conceivably, this extra
degree of accretion freedom in 3-D can produce  a different flux
configuration for scenarios such as in \cite{spr05,dev07}.
\par Perhaps this is what is happening to some extent in KDE. It
is a much longer ergopsheric path to the horizon in terms of
proper distance when $a/M=0.998$ than the KDP simulation,
$a/M=0.9$ (a factor of 2.2) and the in-fall time to the asymptotic
space-time near the horizon is therefore much longer. Initially,
the funnel might fill up as in the $a=0.9M$ case. However, late
arriving flux tubes might have lost so much mass density to the
outflow before they reach the asymptotic space-time near the
horizon that they might actually attain a tenuous plasma state
(i.e., $U_{A}>>1$) before they arrive. As such, there is never
enough inertial force to completely preconditioned the flux tube
for future tenuous approximately force-free inflow. In GHM, the
inertial dynamo forces are generated as the field torques the
plasma on the way to the horizon since the azimuthal twisting is
insufficient in the preconditioned field to allow approximately
force-free inflow \cite{sem04,pun01}. Even if there is an outflow
on these flux tubes, further accretion can still proceed in 3-D by
moving around the flux tubes and a significant GHM effect could
coexist with the high accretion rate the defines quasar activity.
\begin{figure*}
\includegraphics[width=115 mm]{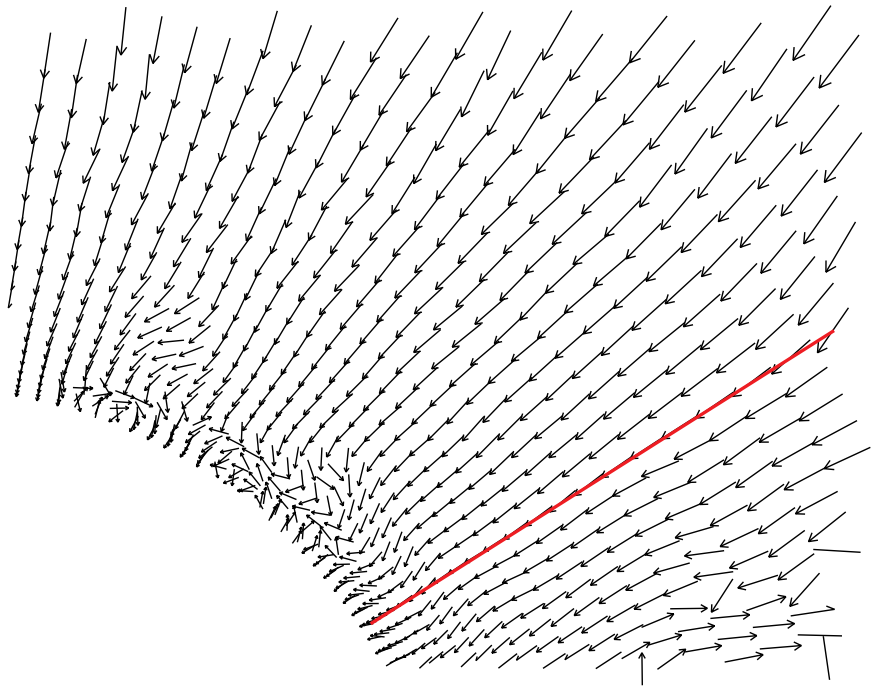}
 \caption{A time snapshot of the $\phi$ averaged poloidal magnetic field
 direction. The arrows indicate direction not magnitude. The
 region is the same as that which is plotted in figure 1. The red line indicates the
 approximate boundary of the funnel. The arrows end
 approximately on a circle that is the inner calculational
 boundary. The plot is provided courtesy of John Hawley.}
    \end{figure*}
\subsection{Non Radial Field Geometry}We verify the near radial field geometry in the funnel
in figure 2 which is a plot of the poloidal field direction
averaged over $\phi$ in a constant time snapshot. The poloidal
field is radial on the boundary surfaces 2 and 4 of the Gaussian
pillbox of figure 1. This does not support the notion of
$S^{\perp}$ flowing along the field lines from the funnel wall, as
discussed in section 3.3, then following abrupt bends of the field
lines toward the radial direction, so as to mimic a source of
radial Poynting flux at $1.2M<r<1.4M$. Near the horizon, at lower
latitudes in the funnel, $50^{\circ}<\theta<55^{\circ}$, where the
strongest $S^{P}$ source is located (the red area in figure 1),
the field is almost exactly radial all the way to the inner
boundary. A small deviation from a radial poloidal field near the
horizon would not be able to produce strong source of $S^{P}$ in
Poynting's theorem. Furthermore, if the poloidal field randomly
oscillates about the radial direction (due to turbulence or
numerical artifacts), near the horizon, then the oscillations in
the Poynting flux would average to zero in the time averaged plot
in figure 1. At higher latitudes in the funnel,
$30^{\circ}<\theta<50^{\circ}$, there is a region of rapid
poloidal field variation right near the inner boundary. The fluid
is clearly highly turbulent, possibly a sign of reflections from
the inner computational boundary. If the strong turbulence is
physical, combined with the $U_{A}>>1$ condition, this would
provide strong magnetic pressure gradients that prevent the
tenuous plasma from flowing force-free into the horizon. This is
the condition described in section 5.3.2 that is most conducive to
a GHM dynamo.
\section{Conclusion} In this paper, we studied simulations of a magnetically dominated
funnel of a rapidly rotating black hole, $a/M=0.998$. A source
region that is responsible for creating over 72\% of $S^{P}$
transported through funnel during the interval $2000M<t<8080M$ of
the simulation was identified. Similarly, one can conclude that
$<28\%$ of $S^{P}$ emerging from the ergospheric funnel is from an
inner boundary source, near the horizon. The small residual
$S^{P}$ injected from the boundary into the accretion wind can be
considered of electrodynamic origin. The distribution of $S^{P}$
in figure 1 is in contrast to the Blandford-Znajek solution in
which essentially all the $S^{P}$ is of electrodynamic origin,
i.e., it emanates from the horizon and passes through the
accretion wind with minimal interaction, thereby maintaining a
virtually constant value along each poloidal flux tube throughout
the ergosphere. The existence of the electromagnetic source within
the simulation was corroborated by existence of a cospatial
electromagnetic angular momentum source using an entirely
different data reduction method. Both of these results follow
directly from the numerical data in \cite{kro05}.
\par There were two possible physical sources for $S^{P}$ in the funnel, there
is the $\mathbf{J}\cdot\mathbf{E}$ source (associated with GHM
inertial currents) and $S^{\perp}$. By time averaging Faraday's
law it was shown that $\mid <F_{\phi \theta}>\mid \gg M\mid
<F_{\phi t}>\mid$. The only source for both $S^{P}$ and
$<T_{\phi}^{\; r}>$ in Poynting's theorem that is consistent with
Faraday,s law was shown to be
$\mathbf{J}\cdot\mathbf{E}=F_{t\theta}J^{\theta}$. It was also
pointed out that the decay of stored field energy, $d\epsilon/dt$
could be the primary source of $S^{P}$. This would suggest that
the simulation is spontaneously creating energy and might be
fatally flawed. In general, there could be a mix of numerical
artifacts, $S^{\perp}$ and $\mathbf{J}\cdot\mathbf{E}$ that
contribute to Poynting's theorem. This can not be sorted out
without further numerical work.
\par  The analysis presented above
suggests some interesting physical possibilities and highlights
the need for further analysis of 3-D simulations about rapidly
rotating black holes ($a/M = 0.998$) in order to clarify the
physics that creates $S^{P}$. Ideally, the 3-D simulation should
be redone in ingoing Kerr-Schild coordinates (to eliminate
boundary reflections a the horizon) and preferably a conservative
code. At present, there is no code efficient enough to handle the
large computing times required. Whether this approach or the
\cite{dev01} approach is used, at least three consecutive time
snapshots are needed in order to find the current distribution
from Ampere's Law at each coarse time step data dump. It is likely
that $J^{\perp}$ is highly turbulent near the event horizon, so it
would be important to then time average the current distribution.
Higher resolution (particularly in the $\theta$ direction) might
be required to resolve the issue of whether the source in figure 1
is from $J^{\perp}E_{\perp}$, $S^{\perp}$ or $d\epsilon/dt$. It
would be prudent, if further calculations in Boyer-Lindquist
coordinates are pursued, to locate the inner calculational
boundary inside of the fast magneto-sonic surface at $r\approx
1.065M$ (inner order to overcome potential reflections). Such
future detailed analysis would determine if there is some new
interesting physics or just some unforseen numerical errors in
KDE.
\par On a more general note, even though the simulations of
\cite{dev03,hir04,dev05,dev06,kro05,gam04,mck05} are very
impressive mathematically for their ability to self consistently
solve a very complicated problem, one must ask if they are
representative of the astrophysical world. It was discussed in
\cite{mck05} that a radial field builds up in the funnel for any
seed field configuration in the torus except for a purely toroidal
field. This implies something generic about the result. The radial
field always grows to seek a pressure balance with the inflow at
the inner edge of the disk, therefore the mass accretion rate sets
the field strength and the power output of the funnel. If this is
a generic property of accretion disks then why are the vast
majority ($\sim 90\%$) of quasars radio quiet. This tight coupling
of the accretion rate to jet power seems to be a serious drawback
of the results. From an astrophysical perspective, it would seem
preferable to develop simulations of scenarios like those
described in \cite{spr05} in order to describe the full panoply of
quasar radio emissivity properties.
\section*{Acknowledgments}I would like to thank
John Hawley and Jean-Pierre DeVilliers for their willingness to
discuss the details of their simulations at great length. I was
also very fortunate that Jonathan Mckinney generously spent time
answering my many questions about his simulations

\end{document}